%% file: DKIST_Cryo_wave_obs.tex
\documentclass[twocolumn]{aastex631}

\usepackage{ifluatex}
\ifluatex
\usepackage{pdftexcmds}
\usepackage{hyperref}
\makeatletter
\let\pdfstrcmp\pdf@strcmp
\let\pdffilemoddate\pdf@filemoddate
\makeatother
\fi
\usepackage{svg}

\usepackage{comment}
\graphicspath{{./}{figures/}}

\usepackage{listings}

\begin{document}

\title{Ubiquitous high-frequency waves and disturbances in the active region corona observed with DKIST/Cryo-NIRSP}

\correspondingauthor{Momchil E. Molnar}
\email{momchil.molnar@swri.org}
\author[0000-0003-0583-0516]{Momchil E. Molnar}
\affiliation{Southwest Research Institute, Boulder, CO}
\affiliation{High Altitude Observatory, 
National Center for Atmospheric Research, Boulder, CO}

\author[0000-0001-5678-9002]{Richard Morton}
\affiliation{Northumbria University, Newcastle upon Tyne, United Kingdom}

\author[0000-0002-3491-1983]{Alin Paraschiv}
\affiliation{National Solar Observatory, Boulder, CO}

\author[0000-0003-0021-9056]{Chris Gilly}
\affiliation{Southwest Research Institute, Boulder, CO}

\author[0000-0002-3699-3134]{Steven R. Cranmer}
\affiliation{Laboratory for Atmospheric and Space Physics, University of Colorado, Boulder, CO, 8030X, USA}

\author[0000-0001-8016-0001]{Kevin Reardon}
\affiliation{National Solar Observatory, Boulder, CO}

\author[0000-0002-7451-9804]{Thomas Schad}
\affiliation{National Solar Observatory, 22 ‘Ohia Ku Street, Pukalani, HI 96768, USA}



\begin{abstract}
The plasma of the solar corona harbors a multitude of coronal wave modes, some of which 
could be dissipated to provide the 
required energy and momentum to heat the corona and accelerate the solar wind. 
We present observations of the corona acquired with the newly commissioned infrared slit spectropolarimeter Cryo-NIRSP at the 
Daniel K. Inouye Solar Telescope (DKIST), Haleakala, Hawaii to study the high frequency wave behavior in closed, active-region structures. Cryo-NIRSP observes the corona off the limb 
in the \ion{Fe}{13} 1074 and 1079\,nm forbidden atomic lines. The large aperture 
of DKIST allows us to explore the active region corona with temporal resolution faster than a second with an achieved spatial 
resolution of 2-5{\arcsec}.
Enhanced wave power is observed in the power spectra up to 100\,mHz. 
Furthermore, we report on a statistically significant anti-correlation 
between the \ion{Fe}{13} 1074\,nm
peak line intensity and line width in our data, possibly pointing to the presence of compressive magnetohydrodynamic (MHD) wave modes. These observations show  
how the powerful spectropolarimetric capabilities of DKIST offer great promise for furthering our knowledge of coronal MHD waves.

\end{abstract}

\keywords{}


\input{Introduction.tex}

\input{Observations.tex}

\input{Observed_PDs.tex}

\input{Line_properties.tex}

\input{Conclusions.tex}
\begin{acknowledgments}

The research reported herein is based in part on data collected with the Daniel K. Inouye Solar Telescope (DKIST) a facility of the National Science Foundation. DKIST is operated by the National Solar Observatory under a cooperative agreement with the Association of Universities for Research in Astronomy, Inc. DKIST is located on land of spiritual and cultural significance to Native Hawaiian people. The use of this important site to further scientific knowledge is done so with appreciation and respect. 
The authors would like to thank Dr. Arpit Shrivastav (SwRI) for the enriching discussions about MHD waves in the solar corona.  
This work was supported in part by NSF Award Number 2408437. The National Center for Atmospheric
Research is a major facility sponsored by the NSF under
Cooperative Agreement No. 1852977. MEM was partly supported through a UCAR/ASP Postdoctoral fellowship during this work.

\end{acknowledgments}

%

\vspace{5mm}


\software{\textit{matplotlib}~\citep{Hunter:2007}, 
          \textit{numpy}~\citep{harris2020array}, 
          \textit{scipy}~\citep{2020SciPy-NMeth},
          \textit{sunpy}~\citep{sunpy_community2020},
          \textit{emcee}~\citep{2013PASP..125..306F},
          cmcrameri colormaps~\citep{zenodo_8409685}.}



\bibliography{DKIST_Cryo_wave_obs}{}
\bibliographystyle{aasjournal}



\end{document}

%% file: Introduction.tex
\section{Introduction} 
\label{sec:intro}

Solar coronal plasma exhibits multiple different wave modes that have been well observed~\citep{2019mwsa.book.....R}. These magnetohydrodynamic (MHD) wave modes
are hypothesized to be the conduits of free energy across the solar atmosphere, sustaining the solar chromosphere and corona at their higher than anticipated temperatures~\citep[e.g., ][]{1999Sci...285..862N,2012RSPTA.370.3217P}.
Observations of coronal waves with imaging instruments have
provided a vast collection of oscillatory phenomena in bright structures well observed on timescales from the tens of minutes down to the tens of seconds~\citep[e.g.,][]{1998ApJ...501L.217D,2002SoPh..207..241P,2011Natur.475..477M,2024A&A...685A..36S}. Compressive waves are usually quickly dampened and are not favored as the leading candidate for wave driven coronal heating. On the other hand, weakly dampened
Alfv\'{e}n waves could possibly carry the required energy up into the corona, where higher-order dissipative processes
dampen those waves through mode conversion, turbulent interactions, and other plasma instabilities and kinetic processes~\citep{2005ApJS..156..265C,2005ApJ...632L..49S,2007ApJS..171..520C}. One of the still open questions about Alfv\'enic wave heating is what contribution these different mechanisms have towards their dissipation. Addressing this requires coronal observations with higher spatial, temporal, and spectral resolution to better constrain the signatures of wave mode conversion and high frequency wave signatures as previous work has suggested the presence of hidden amount of ``dark'' wave energy due to the limited resolution of our observations~\citep{2012ApJ...761..138M,2019ApJ...881...95P}.

The Cryogenic Near Infrared Spectropolarimeter~\citep[Cryo-NIRSP,][]{2023SoPh..298....5F} at the Daniel K. Inouye telescope~\citep[DKIST,][]{2020SoPh..295..172R} is a facility infrared slit spectropolarimeter that can observe both on disk
and as a coronagraphic instrument, allowing for unprecedented off-limb observations with 
better photometric precision
than previous telescopes.
Coronal emission lines like the infrared (IR) \ion{Fe}{13} line pair have been demonstrated to be reliable diagnostic tools for coronal waves~\citep{Tomczyk2009,2020Sci...369..694Y}. Cryo-NIRSP has been successfully commissioned to observe (separately) both forbidden lines of \ion{Fe}{13} at 1074 and 1079\,nm with a spectral resolution 
of \texttt{R}\,$\sim$\,40,000 while covering a wide spectral window around the coronal line, which is essential to constrain the atmospheric
contribution to the signal~\citep{2023ApJ...943...59S}. The large collecting power of the four-meter mirror of DKIST and its
coronagraphic occulter allow Cryo-NIRSP to sample the coronal intensity with a cadence less than a second, as some early results have shown~\citep[e.g.][]{2025ApJ...991...97H, Morton_2025_2}. This makes Cryo-NIRSP an ideal instrument to sample spectroscopically the high-frequency wave regime above 10\,mHz, extending the wave studies with the CoMP instrument~\citep{2008SoPh..247..411T}, and the Norikura coronagraph~\citep{1999PASJ...51..269S}.

This article presents a study of the wave properties observed with Cryo-NIRSP above an 
Active Region (AR) during the Operations Commissioning Phase (OCP2) on 2023 July 6-7. The observations with DKIST were designed to sample
the corona on sub-second timescales while observing in a sit-and-stare mode. The spectroscopy in this work allows for sampling both compressive and Alfv\'enic modes in closed coronal structures. The observations and data reduction are described in Section~\ref{sec:Observations}. The observed propagating disturbances are discussed in Section~\ref{sec:observed_PDs} and the derived power spectra and wave properties are described in Section~\ref{sec:high_freq_behavior}.

%% file: Observations.tex
\section{Observations}
\label{sec:Observations}

\begin{figure*}[ht!]
    \centering
    \includegraphics[width=\textwidth]{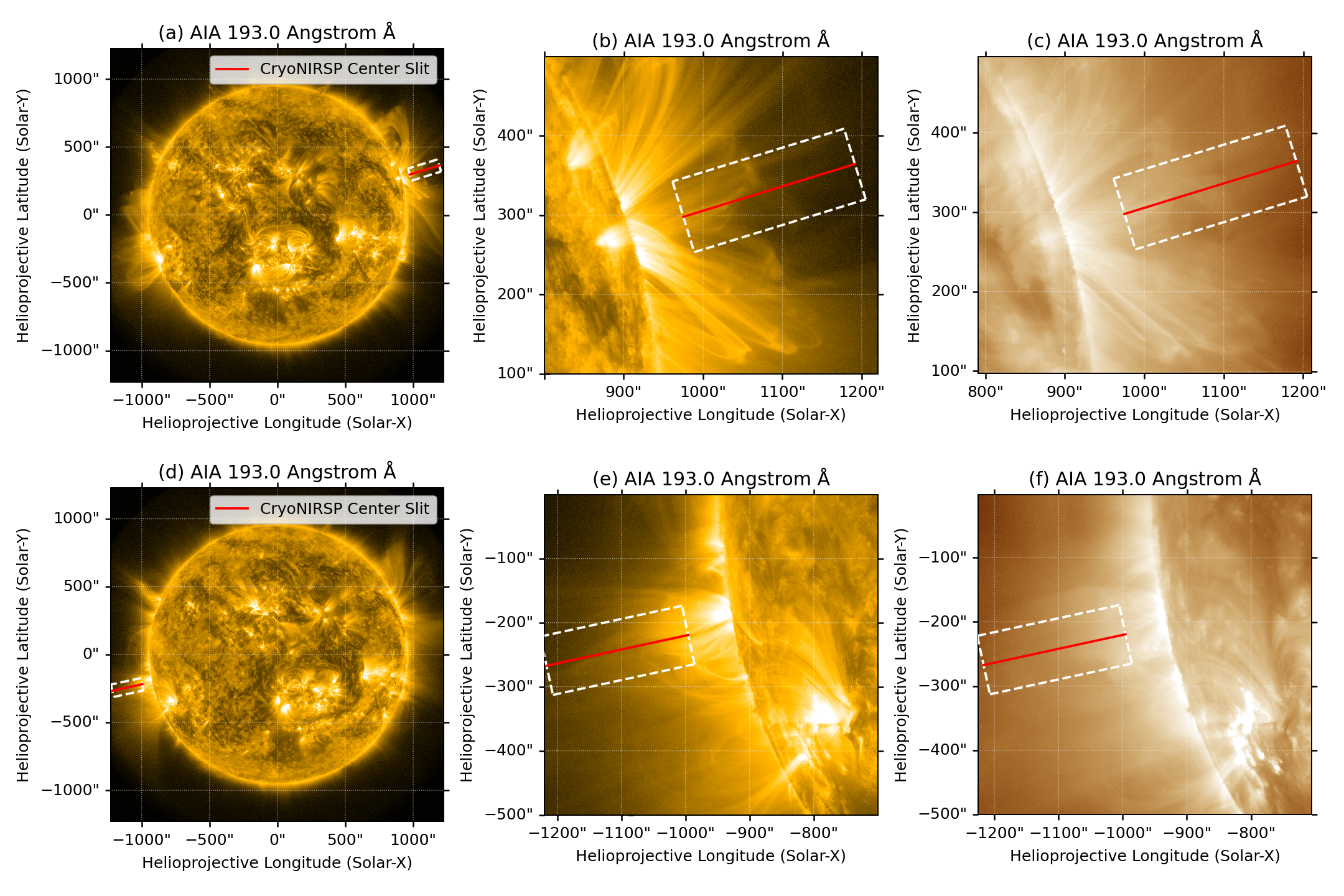}
    
    \caption{Overview from the SDO/AIA 171\,Å and 193\,Å channels of the coronal regions observed with Cryo-NIRSP during the
    observational campaign on July 6-7 2023. The top row, panels (a)-(c) show the 
    observed active region in SDO/AIA channels 171~{\AA} and 193~{\AA} observed
    on July 6 at 19:50 UT. The bottom row follows the same organization of the panels for the observations
    on July 7 2023. The field of view of the Cryo-NIRSP raster is shown as the white rectangle, and the
    sit-and-stare observations are marked as the red line. The magnified subpanels (b), (c), (e), and (f) have RHEF applied. }
    \label{fig:SDO_overview}
\end{figure*}

\begin{figure*}
    \centering
    \includegraphics[width=\linewidth]{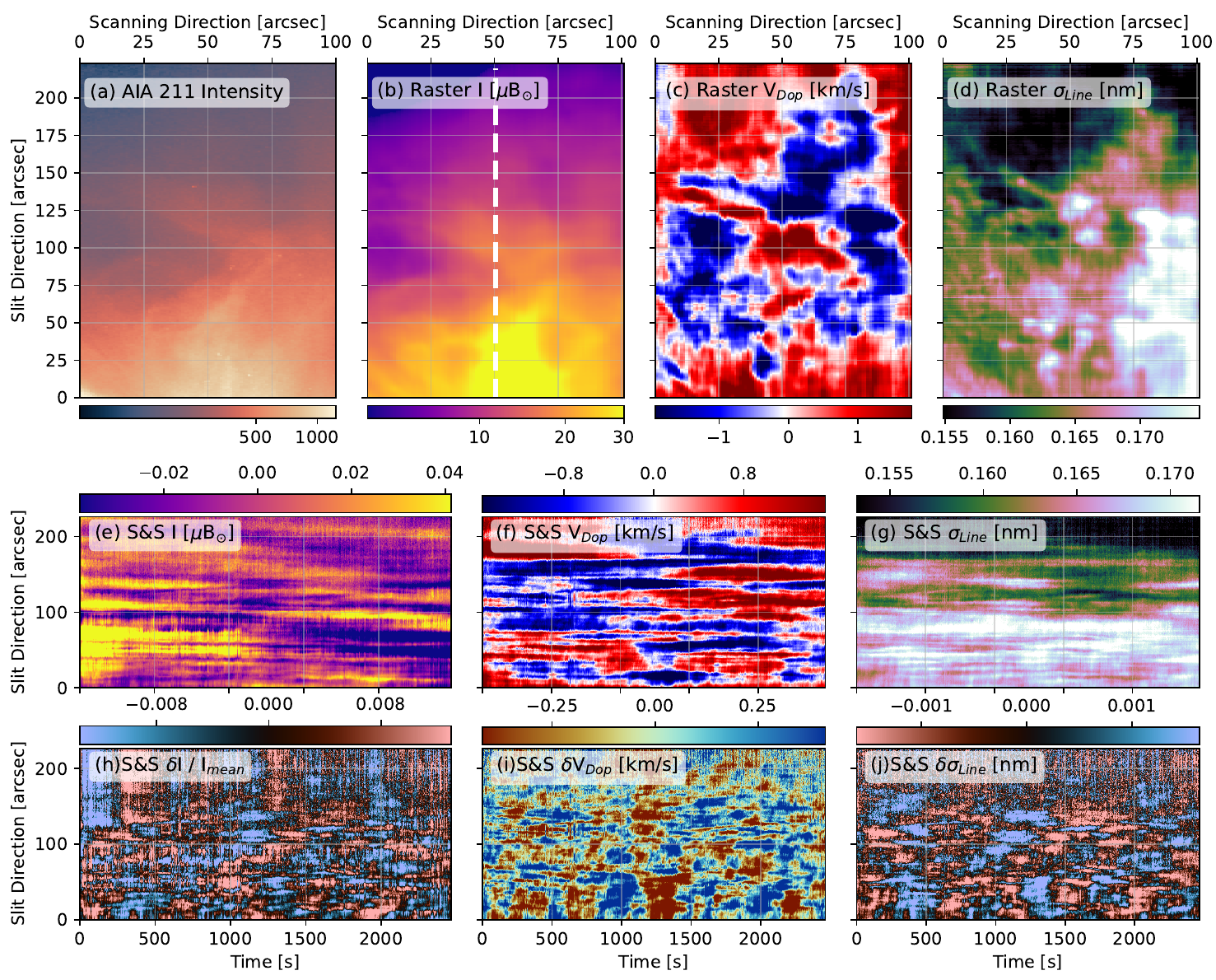}
    \caption{Overview of the DKIST observations for July 6. The top row shows the AIA context image and the wide raster taken with Cryo-NIRSP with the \ion{Fe}{13} 1074\,nm line intensity, Doppler velocity and line width (FWHM in nm) in panels (b)-(d). The corresponding AIA 211 {\AA} image for the DKIST raster is shown in panel (a). The sit-and-stare (S\&S) observations of the same region
    in the second row in panels (e)-(g), where the line properties are shown in the same order. The temporally filtered sit-and-stare data are shown in the bottom row, panels (h)-(j).}
    \label{fig:DKIST_overview_Jul6}
\end{figure*}

\begin{figure*}
    \centering
    \includegraphics[width=\linewidth]{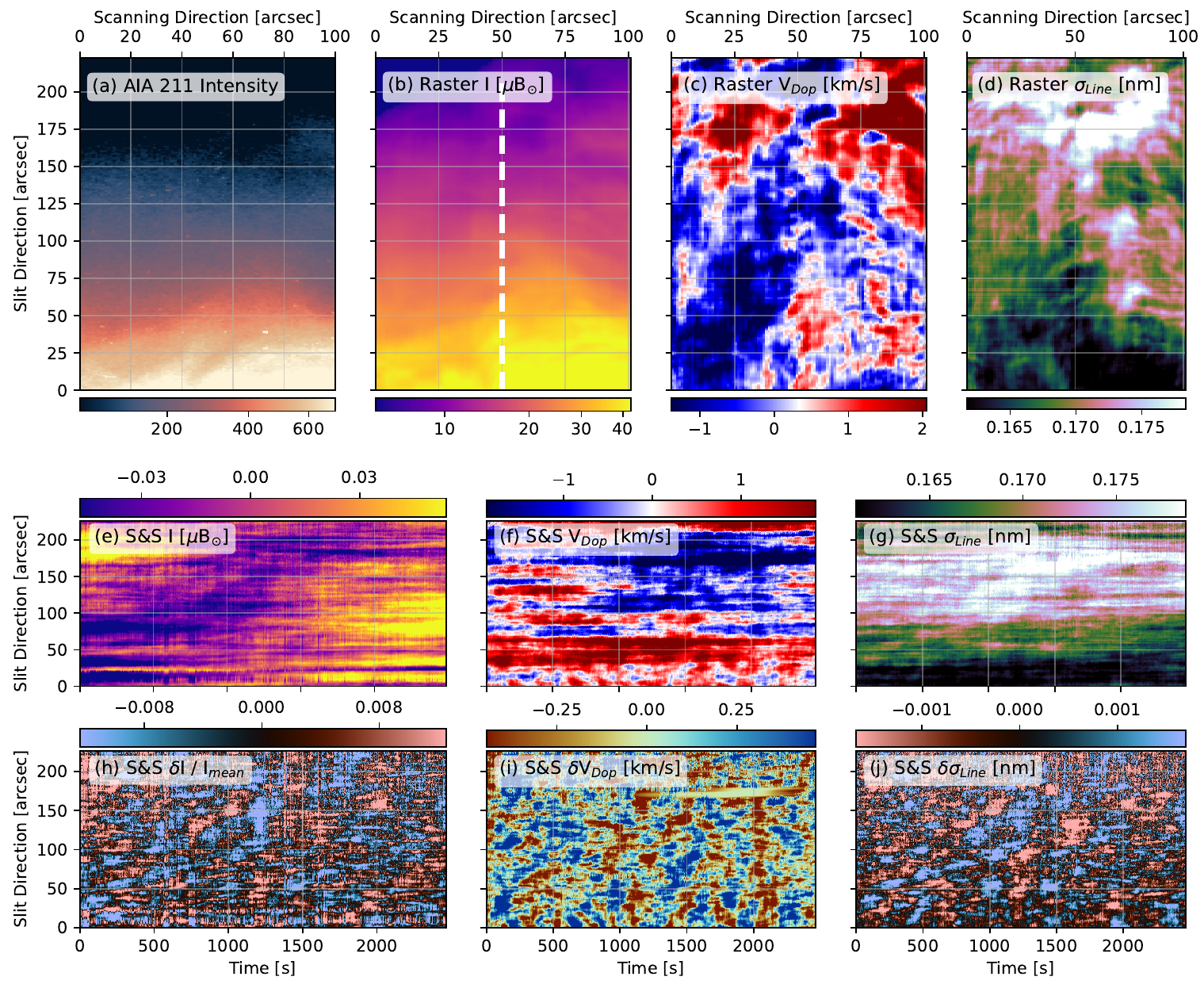}
    \caption{Cryo-NIRSP DKIST observations for July 7 2023, following the same outline as for Figure~\ref{fig:DKIST_overview_Jul6}.  }
    \label{fig:DKIST_overview_Jul7}
\end{figure*}

The DKIST Cryo-NIRSP observations discussed in this paper were taken between 19:06-20:46 UT July 6 and 19:00-20:35 UT July 7 2023 as part 
of observational program PID\,2\_71. During the two observing days,
Cryo-NIRSP observed the corona above active regions NOAA AR 13354 and NOAA AR 13363 
on July 6 and 7 2023, respectively. Cryo-NIRSP was observing with a 0.5\,{\arcsec} wide slit, resulting in effective spectral resolution of R\,$\gtrapprox$\,40000 (obtained from the data analysis in Section~\ref{subsec:Cryo_data_reduction}). In the slit direction, the plate scale of Cryo-NIRSP was
0.12\,{\arcsec}/pixel. 

We utilized two different observing sequences -- a fast sit-and-stare spectroscopic observation for wave study 
and a ``deep'' polarimetric raster scan to provide context of the solar region observed.
The sit-and-stare spectroscopic observations had a radially
 oriented slit and are analyzed to study the high-temporal regime (faster than 1\,Hz sampling)
 of the corona at one fixed slit location. They were designed such that the signal-to-noise ratio (SNR) is sufficient for 
an instantaneous measurement of the spectral line properties along the whole slit. The slit was radially oriented to the solar disc so that propagating
disturbances over the AR can be tracked, with the center of the slit centered at about 1.2\,R$_{\odot}$ (see Table~\ref{table:observations} for exact slit coordinates).
 During the sit-and-stare observations, the \ion{Fe}{13} 1074\,nm spectral line
was recorded with an exposure time of 0.725\,seconds and temporal sampling
 rate of 0.908 seconds for 41.5 minutes, resulting in 2740 frames. The celestial slit extent was
 0.5\,{\arcsec} by 223\,{\arcsec}, corresponding to a projected slit 
length of 164 Mm in the solar corona. To provide context for the sit-and stare observations, 
polarimetric raster scans spanning 223\,{\arcsec} by 101\,{\arcsec} centered around the 
sit-and-stare scans were obtained with Cryo-NIRSP. The raster observations were taken at 101 positions with 
1\,{\arcsec} steps with exposure of 0.725 seconds at 8 modulation states with 4 repeats, which resulted in 51.2 minutes per raster ($\sim$30 seconds per raster position).
The datasets utilized here are all publicly available in the DKIST Data repository, and their metadata and ID's are listed in Table~\ref{table:observations}.

\begin{table*}[ht!]
	\begin{tabular}{c c c c c c}
		\hline
		\hline
		Dataset ID & Date & Time [UT] & Helioprojective &  FOV & Observation type \\
		& & & Coords [arcsec] *& [arcsec] & \\ \hline
		ADMRG & Jul 6 2023 & 19:06:17-19:47:44  & 1078\arcsec, 324\arcsec & 0.5\,{\arcsec} $\times$ 223\,{\arcsec}& sit-and-stare  \\
		ARZLK & Jul 6 2023 & 19:51:01-20:42:10 & 1077\arcsec, 324\arcsec & 101\,{\arcsec} $\times$ 223\,{\arcsec} & raster scan  \\
		BWYRK & Jul 7 2023 & 19:00:21-19:41:48 & -1091\arcsec,-238\arcsec  & 0.5\,{\arcsec} $\times$ 223\,{\arcsec} & sit-and-stare \\
		APGMO & Jul 7 2023 & 19:44:00-20:35:10 & -1091\arcsec, -238\arcsec& 101\,{\arcsec} $\times$ 223\,{\arcsec} & raster scan \\ \hline
	\end{tabular}
	\caption{Summary of the Cryo-NIRSP datasets used in this work. Note that the dataset name in the first column is the universal identifier used in the DKIST data archive. The coordinates for the raster scans correspond to the center of the FOV, and for the sit-and-stare observations to the center of the slit.}
	\label{table:observations}
\end{table*}

Context imaging for the observed regions with Cryo-NIRSP taken from the Solar Dynamics Observatory Atmospheric Imaging Assembly (SDO AIA)~\citep{2012SoPh..275....3P,AIA_paper} is shown in Figure~\ref{fig:SDO_overview}. 
The context images for 2023 July 6 are shown in the top row, and the bottom row shows 
the corresponding 2023 July 7 observations.
The white dashed rectangle marks the region observed by the wide
raster scan and the sit-and-stare sequences were nominally executed at the location of the red line (see alignment discussion below). The zoomed in panels (b), (c), (e), and (f) have been normalized using the Radial Histogram Equalizing Filter (RHEF) \citep{2025arXiv251102798G}. 
Note that
the scanning direction for the rasters is along the narrow side of the raster. As seen 
from the SDO/AIA mages, the Cryo-NIRSP slit intersects distinct coronal loops in the radial direction,
in order to probe the plasma conditions along the loops and search for propagating disturbances.
The top of the slit overlaps with higher overlaying loops and some open field regions, which show
different wave properties demonstrated later in the text. 

\subsection{Reduction of the Cryo-NIRSP data}
\label{subsec:Cryo_data_reduction}

We analyze Cryo-NIRSP Level 1 data obtained from the DKIST Data Center 
Archive, reduced with the current
Cryo-NIRSP pipeline as of writing of this paper. To infer the spectral line properties of the observed \ion{Fe}{13} 1074\,nm line, we follow a reduction procedure outlined in \citet{Schad_2023a} that could be found online hosted on the DKIST community webpage. In summary, this approach is to fit simultaneously models of the instantaneous atmospheric transmission and scattering conditions,
the instrumental response and stray light, the K-corona, and the coronal emission to the data (as shown in \citet{Schad_2023a}) to infer accurately the contribution from the coronal line only. This holistic fitting approach allows for inferring the contribution of the actual solar line,
which is contaminated from multiple different atmospheric and coronal emission sources.  Due to the presence of strong
fringing in the polarimetric measurements, we leave the discussion of the polarimetric part 
of this dataset for future work and only show the resulting intensity rasters for context imaging to support the study of the sit-and-stare observations.

We have aligned the raster images with the corresponding SDO/AIA images in the 171\,{\AA}, 211\,{\AA}, and 193\,{\AA} channels, to assure the careful cospatial coregistration between the different observatories.
We found a $[x, y]$ shift of the DKIST rasters to the SDO/AIA data (assuming SDO/AIA as the ground truth) 
taken on July 6 2024 \emph{ARLZK} dataset of [-5, -7]\,\arcsec, and a shift for the Jul 7 
2024 \emph{APGMO} dataset of [15, 5]\,\arcsec. We have applied these shifts to the sit-and-stare datasets from the same days, assuming the relative shifts are similar to the ones observed in the raster observations. 

We present the resulting spectral line data fits in Figure~\ref{fig:DKIST_overview_Jul6} for the July 6 
2023 data and in Figure~\ref{fig:DKIST_overview_Jul7} for the July 7 2023 data, obtained from the aforementioned inference approach \citep{Schad_2023a}. On the top rows in those figures, a reference AIA 211 image is shown in panel (a), and the \ion{Fe}{13} 1074\,nm line parameters for the scanning raster are shown in panels (b)-(d). The line intensity is displayed in millionths of the center disc brightness $\mu$B$_{\odot}$, calibrated to a close temporal on-disk measurement of flux at the relevant wavelengths (done at the Data Center preprocessing); Doppler 
velocity in km/s; and full width half-maximum (FWHM) in nm are displayed. 
These raster scans correspond to the white dashed rectangles 
in the SDO/AIA overview Figure~\ref{fig:SDO_overview}. For the rest of this manuscript we use the FWHM of the line as its line width. Note that slit direction, coinciding with the radial 
direction away from the solar limb, exhibits strong decrease in the line intensity, not readily
associated with a large scale trend in the FWHM line width or the Doppler velocity. The line intensity and line
widths agree with the previous results from~\cite{Schad_2023a}.

The corresponding sit-and-stare observations are shown in the second row of 
Figure~\ref{fig:DKIST_overview_Jul6} and~\ref{fig:DKIST_overview_Jul7}, where the 
 \ion{Fe}{13} 1074 line fit parameters are shown in the same order as for the raster.
 The location on the sky of the 
 sit-and-stare slit observation is shown as the white dashed line in panel (b) in both figures. In both cases, the sit-and-stare data were taken prior to the raster scans, as discussed in Table~\ref{table:observations}. Due to the large scale trends present on the order of 
 a thousand of seconds, which are not of interest to this study, we remove the long term trends with 
 subtracting the running (local) mean on the order of 675 seconds; at the end of the window we remove the last full averaging window value. The results from this detrending
 are shown in panels (h)-(g) of Figure~\ref{fig:DKIST_overview_Jul6} and  
 Figure~\ref{fig:DKIST_overview_Jul7}. We note that this detrending will remove any signals with 
longer periodicities, but will ensure that the short timescale waves (less than 5 minutes) will be 
enhanced in our data. 
We can see clearly in the line width and intensity 
persistent structures, that will be discussed further in the following sections. Furthermore, there are clear oscillatory trends in all line parameters which
are discussed in detail in Section~\ref{sec:high_freq_behavior}. The fluctuations in the filtered sit-and-stare observations of all three parameters on the scales of tens to hundred seconds exhibit amplitudes significantly lower than the steady background of the quiescent state of the corona, which has been previously observed~\citep{2003ApJ...585..516S}.

%% file: Observed_PDs.tex
\section{Propagating disturbances in the sit-and-stare data}
\label{sec:observed_PDs} 

The most prominent features in the filtered sit-and-stare observations shown in 
Figures~\ref{fig:DKIST_overview_Jul6} and \ref{fig:DKIST_overview_Jul7} are the 
coherent structures found ubiquitously along the slit with length scales of a few tens of Mm and temporal extent of a few hundred seconds. In the case of the Doppler velocity, these might correspond to the Alfv\'enic waves described previously in the literature~\citep{Tomczyk2009, 2016ApJ...828...89M, Morton2025}, but we will further explore their nature by comparing the Doppler velocity with the observed compressive signatures in the other line parameters. We further study the origin of the disturbances found in the line core intensity and line width in the following paragraphs.

 \begin{figure*}[ht!]
	\centering
	\includegraphics[width=\linewidth]{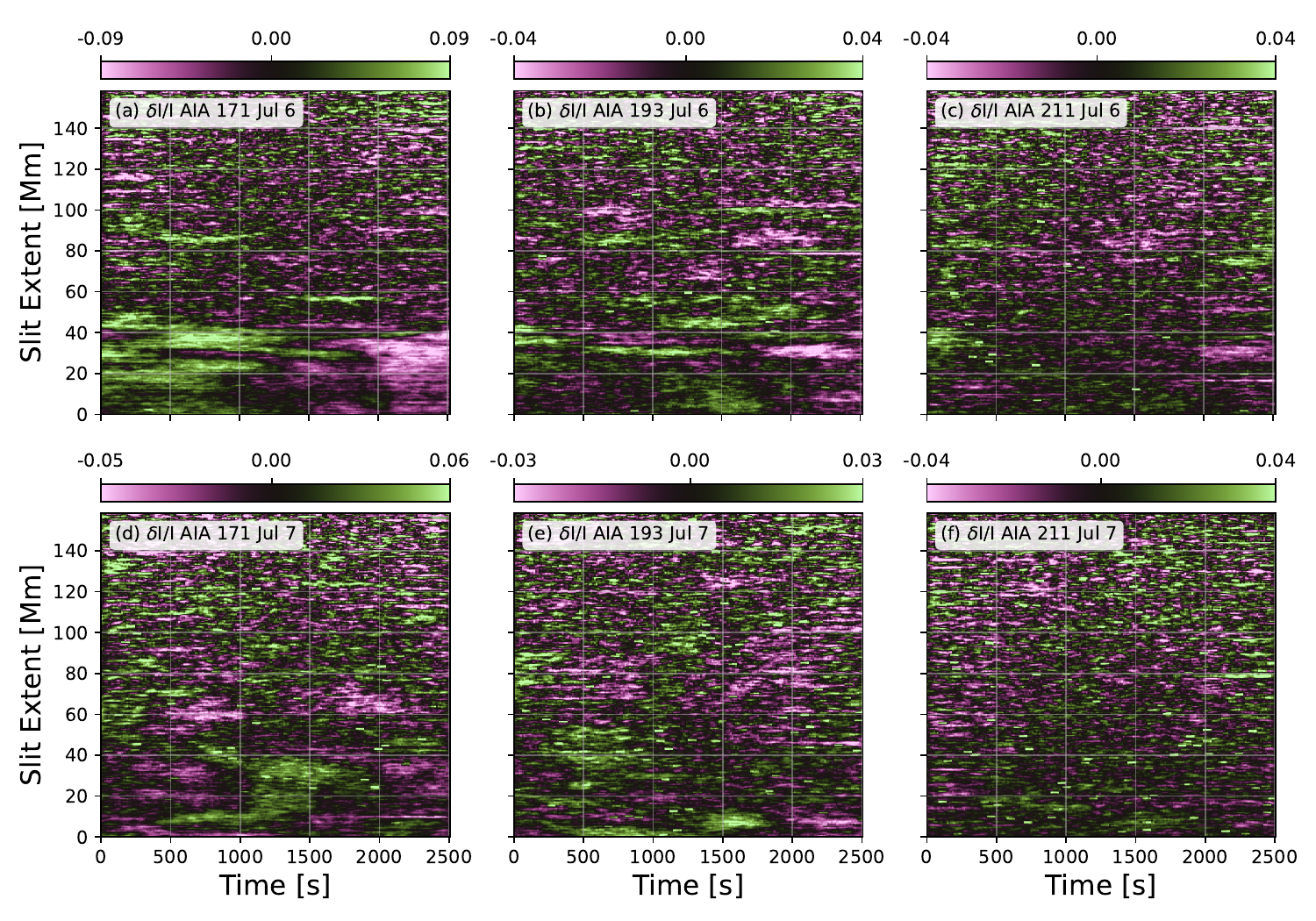}
	\caption{Synthetic AIA Artificial slit images for the corresponding time and position as the CryoNIRSP slit showing the relative intensity fluctuations. The images with removed running average and normalized to the mean intensity are shown at each row, to show the relative intensity fluctuations. Panel (a): July 6, AIA 171\,{\AA} data; panel (b): July 6, 193\,{\AA}; panel (c): July 6, AIA 211\,{\AA};
	panel (d): July 7, AIA 171\,{\AA}; (e) July 7, AIA 193\,{\AA}; (f) July 7, AIA 211\,{\AA}.}
		\label{fig:AIA-artficial-slit}
\end{figure*}

To compare these observations with those obtained from SDO/AIA data, we have computed
synthetic sit-and-stare observations from AIA, shown in Figure~\ref{fig:AIA-artficial-slit}. For this
figure, the corresponding spatiotemporal AIA intensity at the location of the Cryo-NIRSP slit in the 171\,{\AA} and
 193\,{\AA} and 211\,{\AA} channels was retrieved. Following the reduction procedure for the Cryo-NIRSP data
 we removed the running temporal average and then divided the intensity changes by the mean intensity
 over the time frame for each row, so that every column shows the normalized relative intensity fluctuations. We have also binned the AIA data with a 2$\times$2 pixel kernel to decrease further the effects of the noise. The most important result from this comparison is that the fluctuating features seen in the DKIST/Cryo-NIRSP data
 are not clearly detected in the SDO/AIA data. We increased the size of the SDO/AIA averaging kernel to  3$\times$3 to further reduce the noise, 
 but this did not change  the conclusions from Figure~\ref{fig:AIA-artficial-slit}. 
  We believe the non-detection in the AIA
 datasets to be due to the low signal in the UV corona above 1.2\,$R_{\odot}$ where half of the Cryo-NIRSP slit lies. This clearly shows the significantly 
 higher sensitivity to coronal disturbances in the Cryo-NIRSP data, compared to the SDO/AIA observations. 

\begin{figure*}
    \centering
    \includegraphics[width=\linewidth]{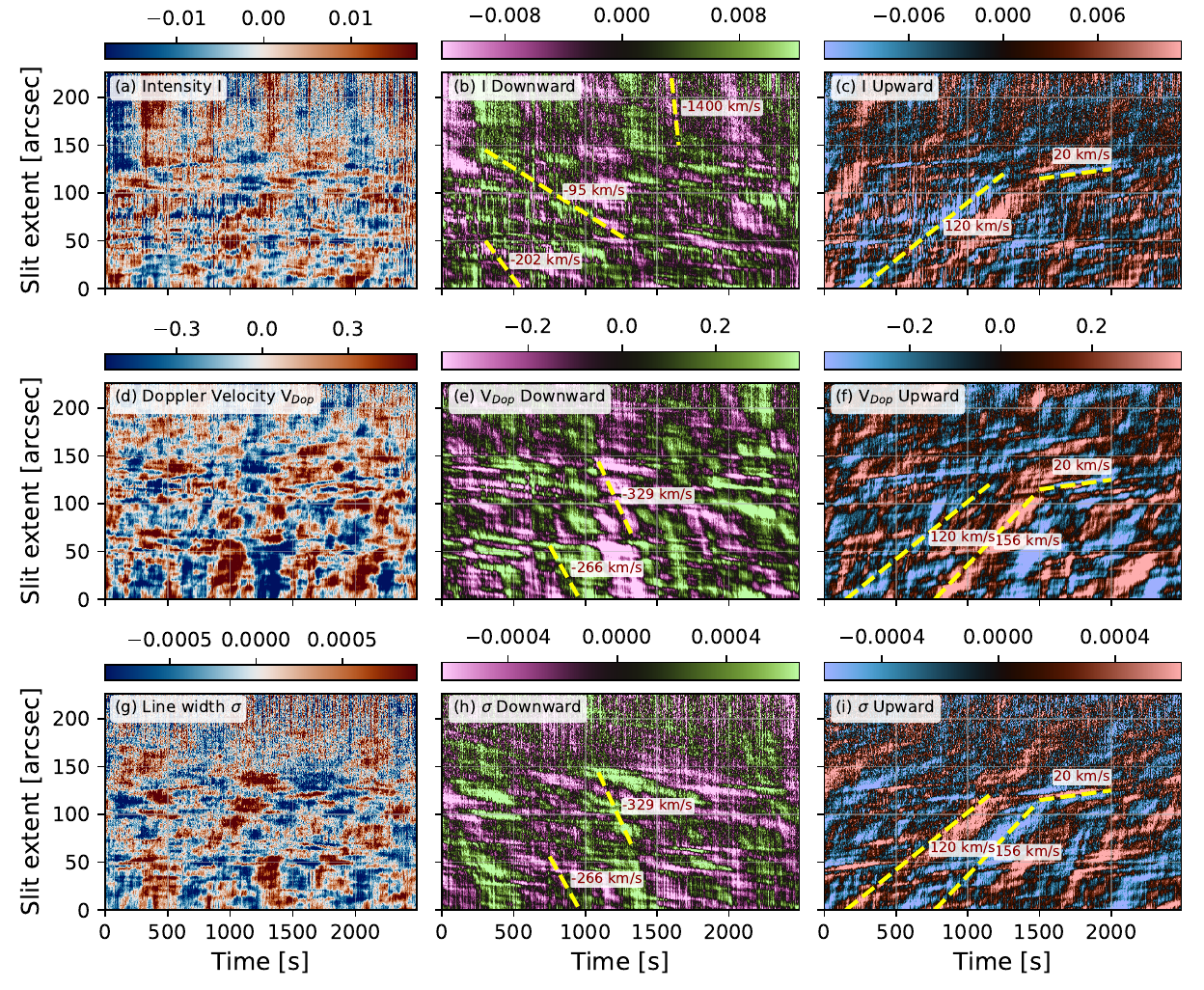}
    \caption{Decomposition of the upward and downward propagating fluctuations of the Cryo-NIRSP sit-and-stare data obtained on Jul 6 2023, based on a $k-\omega$ analysis. Top row: intensity fluctuations $I$;
    middle row: velocity fluctuations V$_{Dop}$; bottom row: line width $\sigma$
    fluctuations. The left column (panels (a), (d),(g)) shows the original data, the middle column the downward (panels (b), (e), and (h)), component, and the right column (panels (c), (f), and (i)) the upward (positive $k$) component.}
    \label{fig:k-omega-decomp_Jul6}
\end{figure*}

To study the properties of the disturbances seen in the filtered sit-and-stare data shown in Figures~\ref{fig:DKIST_overview_Jul6} and \ref{fig:DKIST_overview_Jul7}, we filtered the spectral line property data in $k-\omega$ space to disentangle the upward and downward propagating features~\citep{2014ApJ...787..124D,Tomczyk2009}. $k-\omega$ (wave number-frequency) space is the 2-dimensional Fourier transform of the space-time sit-and-stare observations. We assumed all negative $k$ values to be downward propagating features, and all positive $k$ values with upward propagating features. The results from this computation are shown in Figure~\ref{fig:k-omega-decomp_Jul6} for the observations on July 6, where similar results were obtained for the 
data obtained on July 7.
   
Figure~\ref{fig:k-omega-decomp_Jul6} shows the resulting decomposition of the fluctuation features seen in the Cryo-NIRSP sit-and-stare data obtained on July 6. We have computed those by keeping the positive/negative $k$ values in our $k-\omega$ decomposition and then performed an inverse 2D Fourier transform to create the data in Figure~\ref{fig:k-omega-decomp_Jul6}. The 
left column shows the original sit-and-stare data for the line core intensity (panel (a)), Doppler velocity (panel (d)), and Line width (panel (g)). The filtered downward propagating features are shown in the middle row (panels (b), (e), and (h)), while the upward propagating features are shown in the right column (panels (c), (f), and (i)). We note the following caveat in the following analysis -- due to the fact that we measured the properties with a stationary slit, we can only estimate phase (propagation) speeds along the (radial) direction of the slit. This limits our ability to estimate the actual propagation speeds of the fluctuations and gives us a lower bound of their actual values 
\citep[see further discussion and Figure 1 in][]{Sakurai_2002}.

To note the apparent disturbances in the filtered data, we have manually fitted phase speed 
contours (yellow dashed lines) with the corresponding propagation speeds in Figure~\ref{fig:k-omega-decomp_Jul6}. We note that the propagation speeds of the region associated with the AR (below coordinate y=125\,{\arcsec}) are on the order of a few hundred km/s, where the downward propagating features tend to exhibit significantly faster phase speeds than the upward propagating features, in the realm of a few hundred km/s. These
speeds are significantly lower than the Alfv\'enic wave speeds above ARs~\citep{2005ApJS..156..265C}, and somewhat close to the sound speed in these regions. However, since the slit is observing a fixed coronal location, these estimates are lower bound of the propagation speeds. We also note the significant resemblance between the line intensity (top panel) 
and line width (bottom panel) disturbances, which would be later explored in Section~\ref{subsec:coherence_analysis}.
Also, the behavior of the propagating disturbances change at location of y=140\,{\arcsec}, where the structures in the diagrams become vertical streaks of different durations. If these are solar in origin, they indicate the presences of significantly higher phase speeds. This regions is associated with the location of the slit sampling the overlaying and open magnetic field region above the AR, as shown clearly in Figure~\ref{fig:SDO_overview}. The apparent significantly higher phase speed in this region (about 1,400\,km/s) might be associated with the significantly lower plasma density in these features. These phase speeds correspond well to the local Alfv\'enic speed in the corona~\cite{2004psci.book.....A}. However, we cannot distinguish between wave propagation and mass flows, which will be left for future work. More interestingly, there is a transition region between the two regimes at y=125\,{\arcsec}, which exhibits very
low propagation speeds (V$_{ph}$~20-30\,km/s). This area is clearly associated with the very slanted coronal structure seen at this height in the SDO/AIA\,211 data in  Figure~\ref{fig:DKIST_overview_Jul6} panel (a). 
It is important to note that the structures seen at different locations along the slit may not be spatially connected. Since the coronal plasma is optically thin, we are seeing the projection of structures at different angles, in front of or behind the Sun, all projected onto the plane of the sky which complicates the interpretation of these data.

\begin{table}[]
\resizebox{\columnwidth}{!}{%
\begin{tabular}{llll}
 Day & Region & Line parameter  & \multicolumn{1}{l|}{D/U ratio}   \\ \hline
 July 6 &  AR & Core Intensity & 0.81 (0.01) \\
 &  AR & Doppler Velocity & 0.94 (0.03) \\
 &  AR & Line width & 0.93 (0.03) \\ \hline
  July 6 &  Above AR & Core Intensity & 1.71 (0.09) \\
 &  Above AR & Doppler Velocity & 0.88 (0.09) \\
 &  Above AR & Line width &  1.40 (0.02) \\ \hline
 July 7 &  AR& Core Intensity & 0.60 (0.03) \\
 &  AR & Doppler Velocity & 0.83 (0.02) \\
 & AR & Line width & 0.71 (0.02)
\end{tabular}%
}
\caption{Downward to upward (D/U) propagating disturbance power ratio of propagating disturbances in the Cryo-NIRSP data. The values in parenthesis show the uncertainties.}
\label{table:DU_ratio}
\end{table}

Based on the propagation decomposition described above, we can compute the approximate 
upward- and downward-propagating-disturbance amplitudes seen in the the different spectral line fitting parameters. These measurements provide an estimate of the amount of outgoing to incoming fluctuations and their associated power, which is an useful constraint for coronal heating models~\citep{2019ARA&A..57..157C}. For example, proposed models such as the Parametric Decay Instability (PDI, \cite{1978ApJ...219..700G,2022Univ....8..391M}) or those that utilize a known cascade rate for Alfv\'enic turbulence \citep{1986JGR....91.4111H,1995PhFl....7.2886H}, can effectively dissipate Alfv\'enic waves imprinting 
a signature mixture of daughter waves with a mixture of 
Alfv\'enic and compressive waves propagating forward and backward~\citep{2022JPlPh..88e1501S}. The ratios of the downward to upward fluctuations in the different regions are presented in Table~\ref{table:DU_ratio}. In particular, we have separated for both days the regions of 
the FOV corresponding to the bright AR loops, and the regions of overlying (open) field lines seen at the top of the radial Cryo-NIRSP slit. We have estimated the uncertainty in
the calculated ratios by sampling 50 random slit regions with similar extent, centered on 
the regions studied, and those values are quoted in the parenthesis in Table~\ref{table:DU_ratio}. The resulting ratios in both AR are about 0.8-0.9 for the Doppler velocity and a bit lower for the line intensity and line-widths. Despite those quantities not being directly related to the composition of the wave mode mixture in the solar plasma, they provide a quantitative constraint on any future model of wave mixing in the solar corona. We also investigated the region above the AR observed on July 6 that corresponds to overlaying and open magnetic field regions (not present in the observations taken on July 7th). In this region, outside of the AR plasma, one sees that the ratio of the downward- to upward propagating Doppler velocity fluctuations is about the same, while the fluctuations of the line intensity and line width are significantly higher than unity. This is a signature of the changing plasma wave composition, compared to the AR regions described above. In either case, the close to unity incoming and outgoing waves could be a signature of effectively operating PDI-related heating process operating in the corona~\citep{1978ApJ...219..700G} and these measurements can act as a constraint on future simulations of PDI-based heating scenarios.

The origin of these propagating disturbances is not readily clear, but we believe they are of solar origin. Before turning to the possible coronal explanations, it is crucial to examine the possible instrumental effects creating such large scale features in the data. One possible explanation is 
slight tracking drift of the telescope, which could produce an observed drift of the observed
structures across the slit. When DKIST is pointing above the limb, it is tracking in an open loop mode, relying on the telescope model. Flexure or other subtle changes in the telescope structure could result in a slow drift of the telescope pointing that would not be reflected in the telescope metadata. Cryo-NIRSP has a slitjaw imager that might help characterize image motion on a variety of temporal scales, but it was not properly operational during this phase of the commissioning process. Nonetheless, we believe such drifts are not a viable explanation for the following reasons. First and foremost, 
the lack of coherent drift across the whole FOV in all spectral line parameters is not compatible with this hypothesis. Second, the different propagation speeds across the 
slit, associated with different coronal structures shown above does not support this explanation either. 
We believe these ubiquitous propagating disturbances to be associated with actual coronal 
origin, either of wave origin or mass flows. Furthermore, their imprint in the line core intensity (and line width) point towards a compressive origin. This is not surprising, as fast and slow modes have been extensively observed above ARs~\citep[see review by][]{Nakariakov_ARAA_waves_2020}.

%% file: Line_properties.tex
\section{Observed high-frequency oscillations of the \ion{Fe}{13} 1074}
\label{sec:high_freq_behavior}

\begin{figure*}[htp!]
    \centering
    \includegraphics[width=\linewidth]{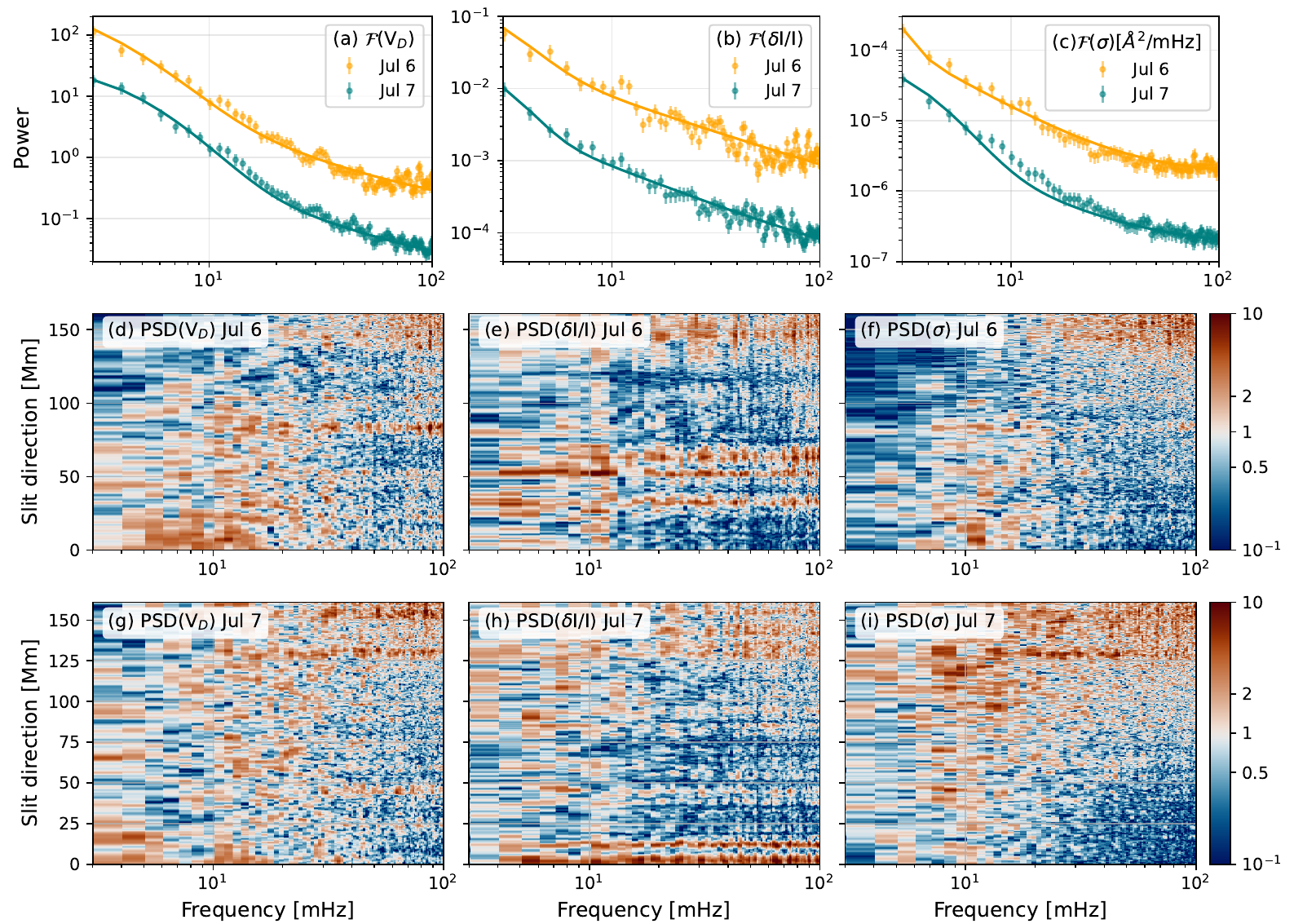}
    \caption{Power spectral density of the observed spectroscopic quantities for July 6 and 7 Cryo-NIRSP data. Panel (a): Averaged over the whole slit PSD of the Doppler velocity, with the data from July 6 shifted upward by a factor of 10 (units of (km/s)$^2$/mHz); panel (b): same as (a) for relative line intensity variations (arbitrary units); panel (c): same as (a) for the line width (units of {\AA}$^2$/mHz). Panels (d)/(e)/(f): excess or deficit of the local PSD compared to the average model of the PSD for the Doppler velocity/Line intensity/line width for the July 6 observations. Panels (g)/(h)/(j): same as the middle row but for the July 7 data.}
    \label{fig:averaged_PSD_line_properties}
\end{figure*}

To study the nature of the high frequency oscillations in the sit-and-stare data we performed Fourier analysis of the data. We have computed the power spectra by taking the discrete Fourier transform on an 
apodized window. We have computed the power spectra by the 
Welch method \citep{Welch_FFT}, using the implementation in the \texttt{scipy} 
library. We have used Hahn windowing with 1100 samples per window, corresponding to 998.8 second wide windows. This analysis results in power spectra with resolution of 0.5\,mHz and Nyquist frequency of 550\,mHz and they are used throughout the rest of this analysis.

\begin{figure*}[htp!]
    \centering
    \includegraphics[width=\linewidth]{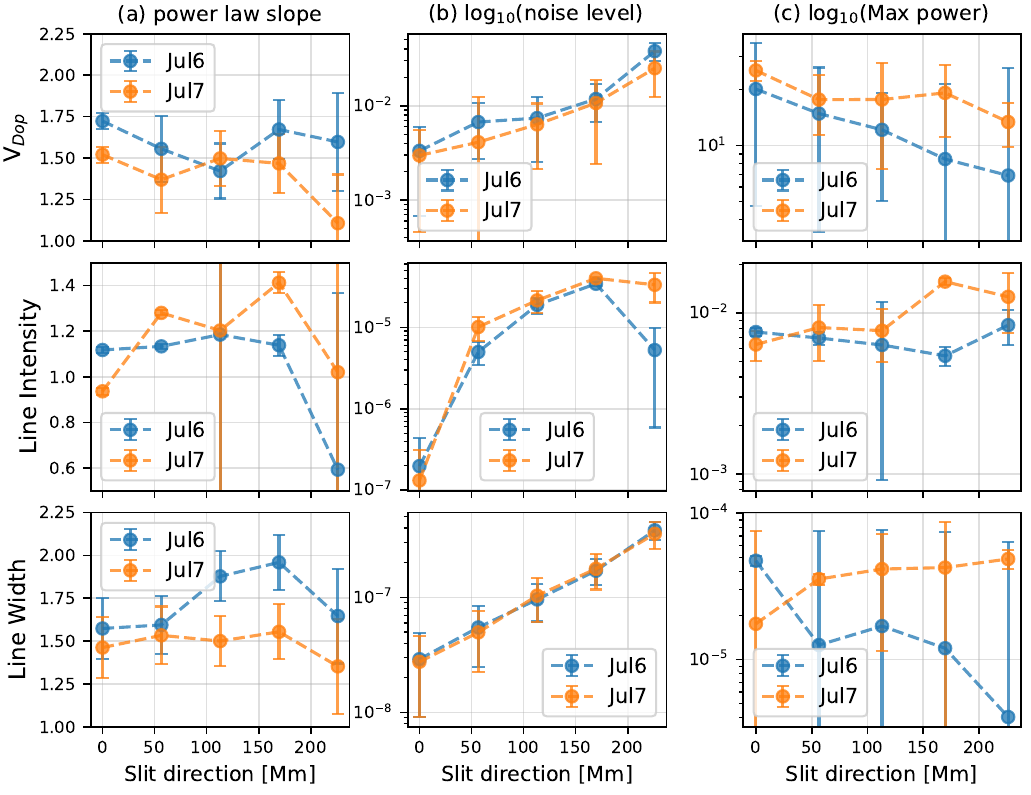}
    \caption{Parameters of the PSDs of the spectral line properties in the sit-and-stare data. Left column corresponds to the power law slopes, middle column to the white noise level, and the right column to the maximum power in the PSD. Top row shows the results for the Doppler Velocity PSD; middle row -- Line core Intensity; and bottom row -- line width.}
    \label{fig:PSD_line_properties}
\end{figure*}

\subsection{High-frequency power in the Cryo-NIRSP data}
\label{subsec:high_frequency_power}

We present the averaged power spectra of the inferred 
detrended sit-and-stare observations in Figure~\ref{fig:averaged_PSD_line_properties}. In the top panels (a)-(c) the averaged power spectra
extending up to 100\,mHz for the spectral line properties are presented. For all three spectral diagnostics 
(Doppler Velocity, relative line intensity fluctuations, and
line width) we find that the white noise floor start dominating the signal at about 50-60\,mHz, extending 
significantly the frequency range of detectable signals compared to the limit of about 10\,mHz with (U)CoMP~\citep{2007Sci...317.1192T, 2016ApJ...828...89M} and the Norikura coronagraph~\citep{1999PASJ...51..269S}.

To model the behavior of the power spectra of the diagnostics observed with Cryo-NIRSP, we perform a maximum likelihood fitting to the power spectra of a model containing a power law, a power enhancement (log-normal form) at around 5\,mHz and a white noise component describing the coronal power spectra (following \cite{2019NatAs...3..223M}): 

\begin{equation}
    \mathcal{F}(f_i) = a_0 \exp{-a_1\times f_i} + a_2 + a_3 \exp{(\log{f_i} - \mu)^2 / \sigma^2} 
\label{eqn:PSD}
\end{equation}

Equation~\ref{eqn:PSD} represents combination of a power law (first term), white noise (second term) and a log-normal excess power  associated with the ubiquitous power found at around 5\,mHz region. We have sampled the posterior distribution of the inferred parameters through a Monte Carlo Markov Chain estimator 
from the \texttt{emcee} package~\citep{2013PASP..125..306F}, imposing similar priors to the one found in \citet{Morton2025}. 

The fits to the average PSDs are shown as the solid lines in Figure~\ref{fig:PSD_line_properties} in panels (a)-(c). We have estimated the uncertainties in the PSD levels following the approach described in 
Appendix A of 
\cite{Morton2025}. In our case, we have taken a correlation length of 8\,Mm for all frequencies, overestimating the uncertainties due to the averaging of our data. There are power enhancements in the observed PSDs
from the fitted model in the high frequency regime. The most notable statistically significant one is  
seen in the case of Doppler velocity at around 30\,mHz on July 6, which has a corresponding, despite more noisy, counterpart in the averaged PSD of the relative intensity fluctuations (panel (b)). There is also a statistically significant excess of power at 10\,mHz in the Doppler velocity signal for both days, which also corresponds to significant enhancement in the relative line intensity and line width fluctuations for the July 6th data. We believe those are signatures of compressive (slow) wave modes above the AR. They correspond to periodicities between 30-100\,s, which have been previously seen with imaging instruments~\citep{1999Sci...285..862N,2018ApJ...868..149K}.

To study the spatial distribution of high-frequency power along the slit, we computed the deviation from the average PSD model for the observed spectral line properties. The resulting ratios of the spectral line parameter PSD to the average model PSD is shown in the bottom two rows of Figure~\ref{fig:averaged_PSD_line_properties}. The columns correspond to the same spectral line parameters, while the middle row correspond to the data from July 6 and the bottom row from July 7. The ratio of the PSDs show significant excess power (red) in regions between 10 and 30\,mHz with spatio-temporal extent pointing towards a coherent source of the signal. These regions of enhanced power seen only in Doppler velocity can be interpreted as Alfv\'enic kink waves~\citep{2007Sci...317.1192T}. However, we also detect significant regions with corresponding power between the Doppler velocity and intensity, which we believe are compressive MHD modes, with periods between 100\,s and 30\,s. We exclude from the discussion the results close to either edge of the slit -- in the case of the bottom of the slit, jitter in the occulter could lead to spurious signals. In the case of the top of the slit, the coronal signals becomes weak and our data suffers from significantly higher white noise, as shown in the next paragraph.

To study the properties of the PSDs of the different spectral line parameters, we computed the average PSD model fit to regions of 340 pixels along the slit direction, resulting in 5 bins along the slit direction. The properties of the PSD models -- power law slope, white noise floor, and total power -- for all spectral line parameters for both July 6 and 8 are shown in Figure~\ref{fig:PSD_line_properties}. Our measurements show a power law slope of the Doppler velocity fluctuations above the AR of about -1.54, which is comparable to previous work~\citep{2016ApJ...828...89M, Morton2025}. Interestingly, this slope for the Doppler Velocity fluctuations in the corona is somewhat lower to the one seen in the Doppler velocity fluctuations in the chromosphere~\citep{2008ApJ...683L.207R, 2021ApJ...920..125M}, which might be due to the physical nature of the oscillations, or the transmissivity of the transition region. More interestingly, the line intensity and line width exhibit power law slopes that do not have significant trends, but have significantly different values -- about -1.2 for the line intensity, and -1.6 for the line width.

Figure~\ref{fig:PSD_line_properties} shows clearly the increase in the white noise floor in all diagnostics with increasing height, which we believe is associated with the decreasing line signal with height. Previous work by \cite{2014ApJ...782L..34D, 2015ApJ...806..273L} have suggested, based on a similar set of observations with UCoMP extending up to 10\,mHz, an increased turbulence towards the loop apices. We believe that our observations with higher sensitivity show clearly that this previous claim is not found in the ARs observed during this campaign. In particular, the combination of the same power law extending up to higher frequencies in our novel observations (past 10\,mHz) and the tight relationship between the decreasing coronal signal away from the solar limb with increasing noise sheds doubt on this previously suggested possibility, as the noise floor in our case is decreased significantly for all diagnostics up to 50\,mHz. However, we do detect a total decrease in the Alfv\'en wave PSD maximal power with increasing height up to the 10\,mHz region as previously claimed, possibly pointing to the decaying nature of fluctuations with increasing height.

\subsection{Coherence of the observed high-frequency coronal fluctuations}
\label{subsec:coherence_analysis}

\begin{figure*}[ht!]
    \centering
    \includegraphics[width=\linewidth]{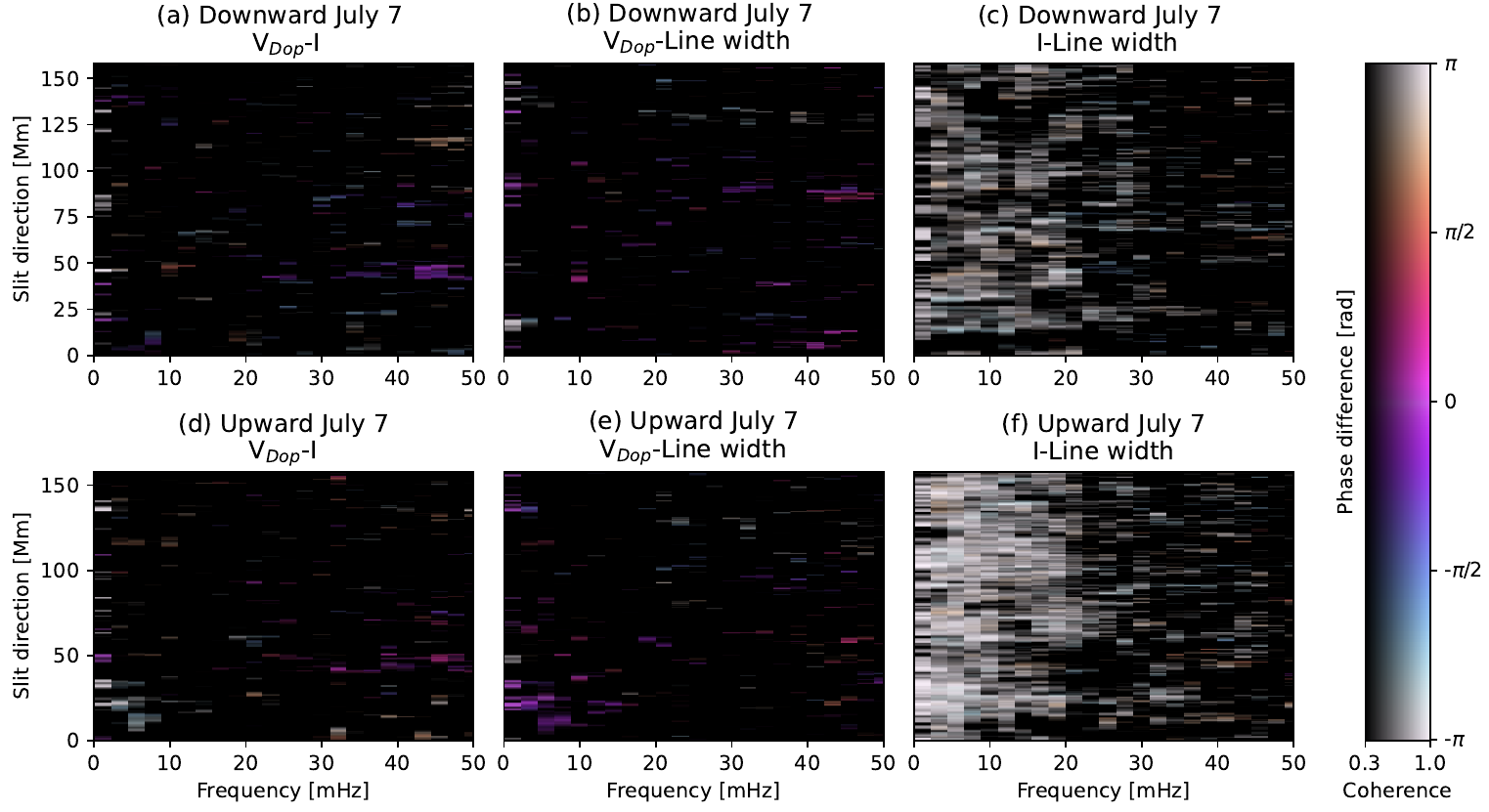}
    \caption{Coherence and phase difference between the different spectral line quantities observed with Cryo-NIRSP for the downward propagating (top row) and upward propagating (bottom row) components of the signals for July 7. Panel (a)/(d): coherence and phase difference between the Doppler velocity and Line Intensity. Panel (b)/(e): coherence between Doppler velocity and Line width; Panel (c)/(f): coherence between intensity and line width for July 6/7.}
    \label{fig:mean_coherency_Jul7}
\end{figure*}

To study the nature of the MHD oscillatory power discussed in the previous section, we use coherence analysis between the different spectral line parameters to infer their nature~\citep[see ][for further description]{2025NRvMP...5...21J}. The coherence $C_{x,y}(f_j)$ at certain frequency $f_j$ of two signals $\left\{x_i, y_i \right\}$ is defined as the absolute value of the cross spectrum of the two signals $\mathcal{F}_{x,y}(f_j)$ normalized by the power spectra of each $\mathcal{F}_{x,x}(f_j)$ and $\mathcal{F}_{y,y}(f_j)$

\begin{equation}
C_{x,y}(f_j) = \frac{|\mathcal{F}_{x,y}(f_j)|^2}{|\mathcal{F}_{x,x}(f_j)||\mathcal{F}_{y,y}(f_j)|}
\label{eqn:coherence}
\end{equation}

The phase difference between the different diagnostics is also computed,
based on the phase difference of the cross power spectrum between the different diagnostics \citep{2025NRvMP...5...21J}. The coherency between the different line diagnostics was practically computed with the Welch method \citep{Welch_FFT} and the 
results are shown in Figure~\ref{fig:mean_coherency_Jul7}. We have estimated the statistically significant level of the coherency between the two signals using the confidence interval based on the 
number of independent segments used in the cross power spectrum computation and the $\chi^2$ distribution nature of the coherency estimates, following the method described in \cite{CoherenceSignificanceLevels}. In our case, the statistical significance level was found to be about 0.31 and all results with values lesser than that are color coded as black in Figure~\ref{fig:mean_coherency_Jul7}.

The results for the whole slit (on the y-axis) for frequencies up to 50\,mHz (along the y-axis) are shown in Figure~\ref{fig:mean_coherency_Jul7}, where a similar results were found for July 6th. We have color coded the lightness of the figure to be reflective of the 
coherence between the two signals (statistically significant signals are non-black), and the color to signify the phase difference between the two signals, with colors corresponding shown in the colorbar. Regions with black color mean that the
coherency there is not statistically significant. We find that 
the only consistently high coherence regions are found between the line intensity and line width fluctuations. In particular, the highest coherence seen is at the low frequency ($<$10\,mHz) regime in the upward propagating signals, in the regions of the 5-minute wave periods, as previously ubiquitously observed by \cite{2019NatAs...3..223M} as Alfv\'enic waves. Interestingly, the phase difference between these line properties is about $\pi$ -- an antiphase relationship. It has been shown that the phase difference between the different plasma parameters in different compressive MHD mode depend on the local plasma parameters and non-adiabatic physical processes in action \citep{2018ApJ...868..149K, 2019PhPl...26h2113Z}. We believe that that this previously unobserved relationship could be used for further study of the coronal plasma conditions and we leave
it for forthcoming work. We note the possibility of potential cross-talk between the inferred line width and intensity that we did not find any evidence in our line fitting approach.

%% file: Conclusions.tex
\section{Conclusions}
\label{sec:Conclusions}

We present coronal spectrographic observations from the DKIST Cryo-NIRSP instrument with less than one second cadence at spatial scales of arcsecond, revealing the characteristics of the high-frequency coronal power spectrum. The observational campaign took place on July 6 and 7 2023 
observing active region loops. The observations took advantage of the DKIST 
capabilities, sampling the corona in the \ion{Fe}{13} 1074\,nm line 
with a cadence of 0.9\,sec and spatial sampling of 
0.12\,{\arcsec}$\times$0.5\,{\arcsec}; however, due to unfavorable seeing conditions the actual resolution of the data could go down to 1-2\,{\arcsec}. We infer the properties of the coronal lines 
after careful processing procedure of the background from sky and F-corona contributions. 

The background change of the sit-and-stare observations is rather slow and 
does not represent a significant small-scale dynamic changes, as shown in the 
non-detrended sit-and-stare data shown in Figure~\ref{fig:DKIST_overview_Jul6}. After 
subtracting the temporal average, we reveal the dynamic nature of the coronal 
structures in the sit-and-stare observations, as shown in 
Figures~\ref{fig:DKIST_overview_Jul6} and \ref{fig:DKIST_overview_Jul7}.

A $k-\omega$ analysis reveals the present upward and downward propagating disturbances, results of which are shown in Figure~\ref{fig:k-omega-decomp_Jul6}. The line intensity and line width seem to exhibit very similar (anticorrelated) behavior, while the 
Doppler velocity shows a different behavior. Furthermore, one can see that the 
different heights in the corona exhibit different outflow and inflow speeds, which correspond to different coronal structures. 
The regions associated with the active region exhibit disturbances with phase speeds on the order of a few hundred kilometers per seconds, which correspond to the acoustic speed in these regions. Hence, we believe that we detect ubiquitous signatures of slow waves in the AR coronal loops, which have been previously detected but not with such abundance, as shown with the comparison with the AIA data in Figure~\ref{fig:AIA-artficial-slit}. The region above the AR loops seen in the July 6 data, the characteristic phase speed is significantly higher, on the order of 1400 km/s, corresponding to the Alfv\'enic speed in the open field corona. We also computed the ratio of the downflowing to upflowing disturbances and found a ratio of lesser than one for all diagnostics. This measurement is related to the dissipation mechanism of Alfv\'enic waves in the corona and further analysis will use it as a basis to constrain the contributions from different Afv\'enic wave mechanisms. 

Examining the power spectra of the line parameters of the sit-and-stare \ion{Fe}{13}
data we see that DKIST can probe an extended regime of the coronal observables, extending previous work with (U)CoMP and the Norikura coronagraphs. The average PSDs exhibit high frequency peaks (excesses) above the 
power law as shown in Figure~\ref{fig:PSD_line_properties}. These high frequency 
excess (between 30 and 100 second periods) 
could be related to MHD waves in the coronal loops, 
with observed periods similar to ones seen before from imaging, but never seen with spectroscopy at such high frequencies. To examine in detail the 
spatial PSD variation, we computed the local excess or depletion of power in the 
PSDs of different diagnostics and the results are presented in the second and third 
rows of Figure~\ref{fig:PSD_line_properties}. We find many regions where
high-frequency oscillatory power is seen localized in particular regions along the 
slit, showing the ubiquitous nature of MHD waves in the AR corona detected with DKIST. Our analysis of the spatial change of the PSD shape with height contradicts the previously suggested idea of enhanced turbulence in coronal loop apices as seen with CoMP~\citep{2015ApJ...806..273L}.

Computing the coherency between the signals found in the different spectral diagnostics 
based on the Welch method show only statistically significant coherence between the line intensity and line width fluctuations, as shown in the results in Figure~\ref{fig:mean_coherency_Jul7}. These fluctuations are associated with almost no Doppler velocity fluctuations.
We find that the mean phase angle between the coherent line width and line 
intensity fluctuations is $\pi$, which corresponds to an antiphase change. This interesting phenomenon was not successfully linked to any instrumental effect; on the other hand, non-zero phases have been predicted for MHD wave properties affected by non-adiabatic plasma processes (and in general changing plasma conditions).

In conclusion, we have reported the coronal power spectrum up to 50 mHz with novel
data from the Cryo-NIRSP instrument, where we have found ample evidence
of ubiquitous fluctuation power at high frequencies, corresponding to MHD waves in the solar corona. We
believe that some of these are associated with previous detections of kink waves detected from (UV) imaging instruments, seen at these frequencies. However, we also detect ever-present compressive waves above the two observed active regions, most probably slow modes, with unexpected phase shift relationship.

%% file: DKIST_Cryo_wave_obs.bbl
\begin{thebibliography}{}
\expandafter\ifx\csname natexlab\endcsname\relax\def\natexlab#1{#1}\fi
\providecommand{\url}[1]{\href{#1}{#1}}
\providecommand{\dodoi}[1]{doi:~\href{http://doi.org/#1}{\nolinkurl{#1}}}
\providecommand{\doeprint}[1]{\href{http://ascl.net/#1}{\nolinkurl{http://ascl.net/#1}}}
\providecommand{\doarXiv}[1]{\href{https://arxiv.org/abs/#1}{\nolinkurl{https://arxiv.org/abs/#1}}}

\bibitem[{{Aschwanden}(2005)}]{2004psci.book.....A}
{Aschwanden}, M.~J. 2005, {Physics of the Solar Corona. An Introduction with
  Problems and Solutions (2nd edition)} ({Springer Nature}),
  \dodoi{10.1007/3-540-30766-4}

\bibitem[{{Crameri}(2023)}]{zenodo_8409685}
{Crameri}, F. 2023, {Scientific Color Maps},  Zenodo,
  \dodoi{10.5281/zenodo.8409685}

\bibitem[{{Cranmer} \& {van Ballegooijen}(2005)}]{2005ApJS..156..265C}
{Cranmer}, S.~R., \& {van Ballegooijen}, A.~A. 2005, \apjs, 156, 265,
  \dodoi{10.1086/426507}

\bibitem[{{Cranmer} {et~al.}(2007){Cranmer}, {van Ballegooijen}, \&
  {Edgar}}]{2007ApJS..171..520C}
{Cranmer}, S.~R., {van Ballegooijen}, A.~A., \& {Edgar}, R.~J. 2007, \apjs,
  171, 520, \dodoi{10.1086/518001}

\bibitem[{{Cranmer} \& {Winebarger}(2019)}]{2019ARA&A..57..157C}
{Cranmer}, S.~R., \& {Winebarger}, A.~R. 2019, \araa, 57, 157,
  \dodoi{10.1146/annurev-astro-091918-104416}

\bibitem[{{De Moortel} {et~al.}(2014){De Moortel}, {McIntosh}, {Threlfall},
  {Bethge}, \& {Liu}}]{2014ApJ...782L..34D}
{De Moortel}, I., {McIntosh}, S.~W., {Threlfall}, J., {Bethge}, C., \& {Liu},
  J. 2014, \apjl, 782, L34, \dodoi{10.1088/2041-8205/782/2/L34}

\bibitem[{{DeForest} \& {Gurman}(1998)}]{1998ApJ...501L.217D}
{DeForest}, C.~E., \& {Gurman}, J.~B. 1998, \apjl, 501, L217,
  \dodoi{10.1086/311460}

\bibitem[{{DeForest} {et~al.}(2014){DeForest}, {Howard}, \&
  {McComas}}]{2014ApJ...787..124D}
{DeForest}, C.~E., {Howard}, T.~A., \& {McComas}, D.~J. 2014, \apj, 787, 124,
  \dodoi{10.1088/0004-637X/787/2/124}

\bibitem[{{Fehlmann} {et~al.}(2023){Fehlmann}, {Kuhn}, {Schad}, {Scholl},
  {Williams}, {Agdinaoay}, {Berst}, {Craig}, {Giebink}, {Goodrich}, {Hnat},
  {James}, {Lockhart}, {Mickey}, {Oswald}, {Puentes}, {Schickling}, {de
  Vanssay}, \& {Warmbier}}]{2023SoPh..298....5F}
{Fehlmann}, A., {Kuhn}, J.~R., {Schad}, T.~A., {et~al.} 2023, \solphys, 298, 5,
  \dodoi{10.1007/s11207-022-02098-y}

\bibitem[{{Foreman-Mackey} {et~al.}(2013){Foreman-Mackey}, {Hogg}, {Lang}, \&
  {Goodman}}]{2013PASP..125..306F}
{Foreman-Mackey}, D., {Hogg}, D.~W., {Lang}, D., \& {Goodman}, J. 2013, \pasp,
  125, 306, \dodoi{10.1086/670067}

\bibitem[{{Gilly} \& {Cranmer}(2025)}]{2025arXiv251102798G}
{Gilly}, C., \& {Cranmer}, S. 2025, arXiv e-prints, arXiv:2511.02798.
\newblock \doarXiv{2511.02798}

\bibitem[{{Goldstein}(1978)}]{1978ApJ...219..700G}
{Goldstein}, M.~L. 1978, \apj, 219, 700, \dodoi{10.1086/155829}

\bibitem[{{Hahn} {et~al.}(2025){Hahn}, {Hofmeister}, {Koukras}, \&
  {Savin}}]{2025ApJ...991...97H}
{Hahn}, M., {Hofmeister}, S.~J., {Koukras}, A., \& {Savin}, D.~W. 2025, \apj,
  991, 97, \dodoi{10.3847/1538-4357/ae017f}

\bibitem[{Harris {et~al.}(2020)Harris, Millman, van~der Walt, Gommers,
  Virtanen, Cournapeau, Wieser, Taylor, Berg, Smith, Kern, Picus, Hoyer, van
  Kerkwijk, Brett, Haldane, del R{\'{i}}o, Wiebe, Peterson,
  G{\'{e}}rard-Marchant, Sheppard, Reddy, Weckesser, Abbasi, Gohlke, \&
  Oliphant}]{harris2020array}
Harris, C.~R., Millman, K.~J., van~der Walt, S.~J., {et~al.} 2020, Nature, 585,
  357, \dodoi{10.1038/s41586-020-2649-2}

\bibitem[{{Hollweg}(1986)}]{1986JGR....91.4111H}
{Hollweg}, J.~V. 1986, \jgr, 91, 4111, \dodoi{10.1029/JA091iA04p04111}

\bibitem[{{Hossain} {et~al.}(1995){Hossain}, {Gray}, {Pontius}, {Matthaeus}, \&
  {Oughton}}]{1995PhFl....7.2886H}
{Hossain}, M., {Gray}, P.~C., {Pontius}, Jr., D.~H., {Matthaeus}, W.~H., \&
  {Oughton}, S. 1995, Physics of Fluids, 7, 2886, \dodoi{10.1063/1.868665}

\bibitem[{Hunter(2007)}]{Hunter:2007}
Hunter, J.~D. 2007, Computing in Science \& Engineering, 9, 90,
  \dodoi{10.1109/MCSE.2007.55}

\bibitem[{{Jafarzadeh} {et~al.}(2025){Jafarzadeh}, {Jess}, {Stangalini},
  {Grant}, {Higham}, {Pessah}, {Keys}, {Belov}, {Calchetti}, {Duckenfield},
  {Fedun}, {Fleck}, {Gafeira}, {Jefferies}, {Khomenko}, {Morton}, {Norton},
  {Rajaguru}, {Schiavo}, {Sharma}, {Silva}, {Solanki}, {Steiner}, {Verth},
  {Vigeesh}, \& {Yadav}}]{2025NRvMP...5...21J}
{Jafarzadeh}, S., {Jess}, D.~B., {Stangalini}, M., {et~al.} 2025, Nature
  Reviews Methods Primers, 5, 21, \dodoi{10.1038/s43586-025-00392-0}

\bibitem[{{Krishna Prasad} {et~al.}(2018){Krishna Prasad}, {Raes}, {Van
  Doorsselaere}, {Magyar}, \& {Jess}}]{2018ApJ...868..149K}
{Krishna Prasad}, S., {Raes}, J.~O., {Van Doorsselaere}, T., {Magyar}, N., \&
  {Jess}, D.~B. 2018, \apj, 868, 149, \dodoi{10.3847/1538-4357/aae9f5}

\bibitem[{{Lemen} {et~al.}(2012){Lemen}, {Title}, {Akin}, {Boerner}, {Chou},
  {Drake}, {Duncan}, {Edwards}, {Friedlaender}, {Heyman}, {Hurlburt}, {Katz},
  {Kushner}, {Levay}, {Lindgren}, {Mathur}, {McFeaters}, {Mitchell}, {Rehse},
  {Schrijver}, {Springer}, {Stern}, {Tarbell}, {Wuelser}, {Wolfson}, {Yanari},
  {Bookbinder}, {Cheimets}, {Caldwell}, {Deluca}, {Gates}, {Golub}, {Park},
  {Podgorski}, {Bush}, {Scherrer}, {Gummin}, {Smith}, {Auker}, {Jerram},
  {Pool}, {Soufli}, {Windt}, {Beardsley}, {Clapp}, {Lang}, \&
  {Waltham}}]{AIA_paper}
{Lemen}, J.~R., {Title}, A.~M., {Akin}, D.~J., {et~al.} 2012, Solar Physics,
  275, 17, \dodoi{10.1007/s11207-011-9776-8}

\bibitem[{{Liu} {et~al.}(2015){Liu}, {McIntosh}, {De Moortel}, \&
  {Wang}}]{2015ApJ...806..273L}
{Liu}, J., {McIntosh}, S.~W., {De Moortel}, I., \& {Wang}, Y. 2015, \apj, 806,
  273, \dodoi{10.1088/0004-637X/806/2/273}

\bibitem[{{Malara} {et~al.}(2022){Malara}, {Primavera}, \&
  {Veltri}}]{2022Univ....8..391M}
{Malara}, F., {Primavera}, L., \& {Veltri}, P. 2022, Universe, 8, 391,
  \dodoi{10.3390/universe8080391}

\bibitem[{{McIntosh} \& {De Pontieu}(2012)}]{2012ApJ...761..138M}
{McIntosh}, S.~W., \& {De Pontieu}, B. 2012, \apj, 761, 138,
  \dodoi{10.1088/0004-637X/761/2/138}

\bibitem[{{McIntosh} {et~al.}(2011){McIntosh}, {de Pontieu}, {Carlsson},
  {Hansteen}, {Boerner}, \& {Goossens}}]{2011Natur.475..477M}
{McIntosh}, S.~W., {de Pontieu}, B., {Carlsson}, M., {et~al.} 2011, \nat, 475,
  477, \dodoi{10.1038/nature10235}

\bibitem[{{Molnar} {et~al.}(2021){Molnar}, {Reardon}, {Cranmer}, {Kowalski},
  {Chai}, \& {Gary}}]{2021ApJ...920..125M}
{Molnar}, M.~E., {Reardon}, K.~P., {Cranmer}, S.~R., {et~al.} 2021,
  Astrophysical Journal, 920, 125, \dodoi{10.3847/1538-4357/ac1515}

\bibitem[{Morton {et~al.}(2025)Morton, Gao, Tajfirouze, Tian, Van~Doorsselaere,
  \& Schad}]{Morton_2025_2}
Morton, R.~J., Gao, Y., Tajfirouze, E., {et~al.} 2025, Nature Astronomy,
  \dodoi{10.1038/s41550-025-02690-9}

\bibitem[{{Morton} {et~al.}(2025){Morton}, {Molnar}, {Cranmer}, \&
  {Schad}}]{Morton2025}
{Morton}, R.~J., {Molnar}, M., {Cranmer}, S.~R., \& {Schad}, T.~A. 2025, \apj,
  982, 104, \dodoi{10.3847/1538-4357/adb8df}

\bibitem[{{Morton} {et~al.}(2016){Morton}, {Tomczyk}, \&
  {Pinto}}]{2016ApJ...828...89M}
{Morton}, R.~J., {Tomczyk}, S., \& {Pinto}, R.~F. 2016, \apj, 828, 89,
  \dodoi{10.3847/0004-637X/828/2/89}

\bibitem[{{Morton} {et~al.}(2019){Morton}, {Weberg}, \&
  {McLaughlin}}]{2019NatAs...3..223M}
{Morton}, R.~J., {Weberg}, M.~J., \& {McLaughlin}, J.~A. 2019, Nature
  Astronomy, 3, 223, \dodoi{10.1038/s41550-018-0668-9}

\bibitem[{Nakariakov \& Kolotkov(2020)}]{Nakariakov_ARAA_waves_2020}
Nakariakov, V.~M., \& Kolotkov, D.~Y. 2020, Annual Review of Astronomy and
  Astrophysics, 58, 441,
  \dodoi{https://doi.org/10.1146/annurev-astro-032320-042940}

\bibitem[{{Nakariakov} {et~al.}(1999){Nakariakov}, {Ofman}, {Deluca},
  {Roberts}, \& {Davila}}]{1999Sci...285..862N}
{Nakariakov}, V.~M., {Ofman}, L., {Deluca}, E.~E., {Roberts}, B., \& {Davila},
  J.~M. 1999, Science, 285, 862, \dodoi{10.1126/science.285.5429.862}

\bibitem[{{Pant} {et~al.}(2019){Pant}, {Magyar}, {Van Doorsselaere}, \&
  {Morton}}]{2019ApJ...881...95P}
{Pant}, V., {Magyar}, N., {Van Doorsselaere}, T., \& {Morton}, R.~J. 2019,
  \apj, 881, 95, \dodoi{10.3847/1538-4357/ab2da3}

\bibitem[{{Parnell} \& {De Moortel}(2012)}]{2012RSPTA.370.3217P}
{Parnell}, C.~E., \& {De Moortel}, I. 2012, Philosophical Transactions of the
  Royal Society of London Series A, 370, 3217, \dodoi{10.1098/rsta.2012.0113}

\bibitem[{{Pasachoff} {et~al.}(2002){Pasachoff}, {Babcock}, {Russell}, \&
  {Seaton}}]{2002SoPh..207..241P}
{Pasachoff}, J.~M., {Babcock}, B.~A., {Russell}, K.~D., \& {Seaton}, D.~B.
  2002, SoPh, 207, 241, \dodoi{10.1023/A:1016297800478}

\bibitem[{{Pesnell} {et~al.}(2012){Pesnell}, {Thompson}, \&
  {Chamberlin}}]{2012SoPh..275....3P}
{Pesnell}, W.~D., {Thompson}, B.~J., \& {Chamberlin}, P.~C. 2012, Solar
  Physics, 275, 3, \dodoi{10.1007/s11207-011-9841-3}

\bibitem[{{Reardon} {et~al.}(2008){Reardon}, {Lepreti}, {Carbone}, \&
  {Vecchio}}]{2008ApJ...683L.207R}
{Reardon}, K.~P., {Lepreti}, F., {Carbone}, V., \& {Vecchio}, A. 2008, \apjl,
  683, L207, \dodoi{10.1086/591790}

\bibitem[{{Rimmele} {et~al.}(2020){Rimmele}, {Warner}, {Keil}, {Goode},
  {Kn{\"o}lker}, {Kuhn}, {Rosner}, {McMullin}, {Casini}, {Lin}, {W{\"o}ger},
  {von der L{\"u}he}, {Tritschler}, {Davey}, {de Wijn}, {Elmore}, {Fehlmann},
  {Harrington}, {Jaeggli}, {Rast}, {Schad}, {Schmidt}, {Mathioudakis},
  {Mickey}, {Anan}, {Beck}, {Marshall}, {Jeffers}, {Oschmann}, {Beard},
  {Berst}, {Cowan}, {Craig}, {Cross}, {Cummings}, {Donnelly}, {de Vanssay},
  {Eigenbrot}, {Ferayorni}, {Foster}, {Galapon}, {Gedrites}, {Gonzales},
  {Goodrich}, {Gregory}, {Guzman}, {Guzzo}, {Hegwer}, {Hubbard}, {Hubbard},
  {Johansson}, {Johnson}, {Liang}, {Liang}, {McQuillen}, {Mayer}, {Newman},
  {Onodera}, {Phelps}, {Puentes}, {Richards}, {Rimmele}, {Sekulic}, {Shimko},
  {Simison}, {Smith}, {Starman}, {Sueoka}, {Summers}, {Szabo}, {Szabo},
  {Wampler}, {Williams}, \& {White}}]{2020SoPh..295..172R}
{Rimmele}, T.~R., {Warner}, M., {Keil}, S.~L., {et~al.} 2020, \solphys, 295,
  172, \dodoi{10.1007/s11207-020-01736-7}

\bibitem[{{Roberts}(2019)}]{2019mwsa.book.....R}
{Roberts}, B. 2019, {MHD waves in the solar atmosphere}

\bibitem[{{Sakurai} {et~al.}(2002){Sakurai}, {Ichimoto}, {Raju}, \&
  {Singh}}]{Sakurai_2002}
{Sakurai}, T., {Ichimoto}, K., {Raju}, K.~P., \& {Singh}, J. 2002, \solphys,
  209, 265, \dodoi{10.1023/A:1021297313448}

\bibitem[{{Schad} {et~al.}(2023{\natexlab{a}}){Schad}, {Kuhn}, {Fehlmann},
  {Scholl}, {Harrington}, {Rimmele}, \& {Tritschler}}]{2023ApJ...943...59S}
{Schad}, T.~A., {Kuhn}, J.~R., {Fehlmann}, A., {et~al.} 2023{\natexlab{a}},
  \apj, 943, 59, \dodoi{10.3847/1538-4357/acabbd}

\bibitem[{{Schad} {et~al.}(2023{\natexlab{b}}){Schad}, {Kuhn}, {Fehlmann},
  {Scholl}, {Harrington}, {Rimmele}, \& {Tritschler}}]{Schad_2023a}
---. 2023{\natexlab{b}}, \apj, 943, 59, \dodoi{10.3847/1538-4357/acabbd}

\bibitem[{{Schekochihin}(2022)}]{2022JPlPh..88e1501S}
{Schekochihin}, A.~A. 2022, Journal of Plasma Physics, 88, 155880501,
  \dodoi{10.1017/S0022377822000721}

\bibitem[{{Shrivastav} {et~al.}(2024){Shrivastav}, {Pant}, {Berghmans},
  {Zhukov}, {Van Doorsselaere}, {Petrova}, {Banerjee}, {Lim}, \&
  {Verbeeck}}]{2024A&A...685A..36S}
{Shrivastav}, A.~K., {Pant}, V., {Berghmans}, D., {et~al.} 2024, \aap, 685,
  A36, \dodoi{10.1051/0004-6361/202346670}

\bibitem[{{Singh} {et~al.}(1999){Singh}, {Ichimoto}, {Imai}, {Sakurai}, \&
  {Takeda}}]{1999PASJ...51..269S}
{Singh}, J., {Ichimoto}, K., {Imai}, H., {Sakurai}, T., \& {Takeda}, A. 1999,
  \pasj, 51, 269, \dodoi{10.1093/pasj/51.2.269}

\bibitem[{{Singh} {et~al.}(2003){Singh}, {Ichimoto}, {Sakurai}, \&
  {Muneer}}]{2003ApJ...585..516S}
{Singh}, J., {Ichimoto}, K., {Sakurai}, T., \& {Muneer}, S. 2003, \apj, 585,
  516, \dodoi{10.1086/346000}

\bibitem[{{Suzuki} \& {Inutsuka}(2005)}]{2005ApJ...632L..49S}
{Suzuki}, T.~K., \& {Inutsuka}, S.-i. 2005, \apjl, 632, L49,
  \dodoi{10.1086/497536}

\bibitem[{{The SunPy Community} {et~al.}(2020){The SunPy Community}, Barnes,
  Bobra, Christe, Freij, Hayes, Ireland, Mumford, Perez-Suarez, Ryan, Shih,
  Chanda, Glogowski, Hewett, Hughitt, Hill, Hiware, Inglis, Kirk, Konge, Mason,
  Maloney, Murray, Panda, Park, Pereira, Reardon, Savage, Sipőcz, Stansby,
  Jain, Taylor, Yadav, Rajul, \& Dang}]{sunpy_community2020}
{The SunPy Community}, Barnes, W.~T., Bobra, M.~G., {et~al.} 2020, The
  Astrophysical Journal, 890, 68, \dodoi{10.3847/1538-4357/ab4f7a}

\bibitem[{Thompson(1979)}]{CoherenceSignificanceLevels}
Thompson, R. O. R.~Y. 1979, Journal of Atmospheric Sciences, 36, 2020 ,
  \dodoi{10.1175/1520-0469(1979)036<2020:CSL>2.0.CO;2}

\bibitem[{{Tomczyk} \& {McIntosh}(2009)}]{Tomczyk2009}
{Tomczyk}, S., \& {McIntosh}, S.~W. 2009, \apj, 697, 1384,
  \dodoi{10.1088/0004-637X/697/2/1384}

\bibitem[{{Tomczyk} {et~al.}(2007){Tomczyk}, {McIntosh}, {Keil}, {Judge},
  {Schad}, {Seeley}, \& {Edmondson}}]{2007Sci...317.1192T}
{Tomczyk}, S., {McIntosh}, S.~W., {Keil}, S.~L., {et~al.} 2007, Science, 317,
  1192, \dodoi{10.1126/science.1143304}

\bibitem[{{Tomczyk} {et~al.}(2008){Tomczyk}, {Card}, {Darnell}, {Elmore},
  {Lull}, {Nelson}, {Streander}, {Burkepile}, {Casini}, \&
  {Judge}}]{2008SoPh..247..411T}
{Tomczyk}, S., {Card}, G.~L., {Darnell}, T., {et~al.} 2008, \solphys, 247, 411,
  \dodoi{10.1007/s11207-007-9103-6}

\bibitem[{Virtanen {et~al.}(2020)Virtanen, Gommers, Oliphant, Haberland, Reddy,
  Cournapeau, Burovski, Peterson, Weckesser, Bright, {van der Walt}, Brett,
  Wilson, Millman, Mayorov, Nelson, Jones, Kern, Larson, Carey, Polat, Feng,
  Moore, {VanderPlas}, Laxalde, Perktold, Cimrman, Henriksen, Quintero, Harris,
  Archibald, Ribeiro, Pedregosa, {van Mulbregt}, \& {SciPy 1.0
  Contributors}}]{2020SciPy-NMeth}
Virtanen, P., Gommers, R., Oliphant, T.~E., {et~al.} 2020, Nature Methods, 17,
  261, \dodoi{10.1038/s41592-019-0686-2}

\bibitem[{Welch(1967)}]{Welch_FFT}
Welch, P. 1967, IEEE Transactions on Audio and Electroacoustics, 15, 70,
  \dodoi{10.1109/TAU.1967.1161901}

\bibitem[{{Yang} {et~al.}(2020){Yang}, {Bethge}, {Tian}, {Tomczyk}, {Morton},
  {Del Zanna}, {McIntosh}, {Karak}, {Gibson}, {Samanta}, {He}, {Chen}, \&
  {Wang}}]{2020Sci...369..694Y}
{Yang}, Z., {Bethge}, C., {Tian}, H., {et~al.} 2020, Science, 369, 694,
  \dodoi{10.1126/science.abb4462}

\bibitem[{{Zavershinskii} {et~al.}(2019){Zavershinskii}, {Kolotkov},
  {Nakariakov}, {Molevich}, \& {Ryashchikov}}]{2019PhPl...26h2113Z}
{Zavershinskii}, D.~I., {Kolotkov}, D.~Y., {Nakariakov}, V.~M., {Molevich},
  N.~E., \& {Ryashchikov}, D.~S. 2019, Physics of Plasmas, 26, 082113,
  \dodoi{10.1063/1.5115224}

\end{thebibliography}
